\documentclass[twocolumn,showpacs,preprintnumbers,amsmath,amssymb,nofootinbib]{revtex4-1}

\usepackage{graphicx}
\usepackage{dcolumn}
\usepackage{bm}

\begin{document}

\title{Flashing flexodomains and electroconvection rolls in a nematic liquid crystal}

\author{P\'{e}ter Salamon$^1$, N\'{a}ndor \'{E}ber$^1$, Alexei Krekhov$^2$ and \'{A}gnes Buka$^1$ }

\affiliation{$^1$ Institute for Solid State Physics and
Optics, Wigner Research Centre for Physics, Hungarian Academy of Sciences, H-1525 Budapest, P.O.B.49,
Hungary}

\affiliation{$^2$ Institute of Physics, University of Bayreuth,
D-95440 Bayreuth, Germany}

\date{\today}

\begin{abstract}
Pattern forming instabilities induced by
ultralow frequency sinusoidal voltages were studied in a rod-like
nematic liquid crystal by microscopic observations and
simultaneous electric current measurements. Two pattern
morphologies, electroconvection (EC) and flexodomains (FD), were
distinguished; both appearing as time separated flashes within
each half period of driving. A correlation was found between the
time instants of the EC flashes and that of the nonlinear current
response. The voltage dependence of the pattern contrast $C(U)$
for EC has a different character than that for the FD. The
flattening of $C(U)$ at reducing the frequency was described in
terms of an imperfect bifurcation model. Analysing the threshold
characteristics of FD the temperature dependence of the difference
$|e_1-e_3|$ of the flexoelectric coefficients were also
determined by considering elastic anisotropy.
\end{abstract}

\pacs{61.30.Gd, 47.54.-r }

\maketitle
\section{\label{sec:intro}Introduction}

Nematic liquid crystals are the simplest paradigm for anisotropic
fluids; i.e. liquids with a preferred direction of the orientation of molecules with anisotropic shape which is described
by the director field $\mathbf{n}$. The anisotropy of their
dielectric properties allows controlling the director by electric
fields. The (usually homogeneous) reorientation of the director by
a properly applied voltage changes the direction of the optical
axis and hence the light transmittance of the sample; this forms
the physical background of the liquid crystal displays,
\cite{Blinov-book} used widespread in common electronic devices.

Applying an electric voltage to a nematic liquid crystal layer
can, however, often result in the appearance of spatio-temporal,
periodic or disordered structures too.  The conditions of their
occurrence, the pattern morphologies and their onset
characteristics have been extensively studied since decades, both
experimentally and theoretically \cite{Blinov-book,Bobylev1977,Buka2006,Buka2007,Bodenschatz1988,Krekhov2008,
Treiber1995,May2008,Eber2012,Krekhov2011,Kramer1996,
Kramer2001,Kochowska2004,Buka2012}.

In the mostly studied planar configuration, where the director is
initially oriented parallel to the confining plates, one of the
electric field induced patterns corresponds to spatially periodic,
equilibrium director deformations (seen as stripes parallel to the
director in a polarizing microscope), occurring due to a
flexoelectric free energy gain of the deformed state; therefore
they have been coined \emph{flexoelectric domains} (FDs)
\cite{Bobylev1977}. FDs have so far been detected in a few nematic
compounds only and they are observable at DC (or very low
frequency AC) driving only.

A more frequent, but also more complex pattern forming phenomenon
is the electroconvection (EC) where the director distortions are
accompanied by space charge separation and hence by material flow;
thus having a dissipative character. It could be observed in many
nematics, some of which possess substantially different material
properties \cite{Buka2006,Buka2007}. EC patterns could be induced
in a wide frequency range of the applied voltage (ranging from DC
up to several hundreds kHz AC); the resulting convection rolls are
seen in a polarizing microscope as stripes whose direction may be
normal to, oblique or parallel with the director. Up to now
studies were mostly focussed on the class of nematics with
negative dielectric and positive conductivity anisotropies and on
driving frequencies $f$ within the range of 10 Hz to 10 kHz. In
this $f$ range evolution of the pattern requires numerous driving
periods after voltage application. For such conditions the
variation of pattern morphologies (conductive and dielectric
regimes, oblique and normal rolls) upon the amplitude and
frequency of the applied voltage have been explored in detail and
the mechanism as an electrohydrodynamic instability has been well
understood. A quantitative theoretical description of the pattern
threshold, the critical wave vector and some secondary transitions
(e.g. abnormal rolls) could be given combining nematodynamics with
electrodynamics under the simplifying assumption of Ohmic
conductivity -- now called as the standard model of EC
\cite{Bodenschatz1988} -- or via its extensions by
flexoelectricity \cite{Krekhov2008} or by ionic
diffusion/recombination \cite{Treiber1995}.

Recently interest has arisen to study the behaviour in another,
subhertz frequency range, where the pattern growth/decay times are
(much) shorter than the driving period, using compounds which may
exhibit both EC and FD patterns. It has been proven experimentally
that at such ultralow frequencies both for the dielectric
\cite{May2008} and the conductive \cite{Eber2012} EC regimes, as
well as for the FD \cite{May2008,Eber2012} the patterns are
flashing, i.e. they exist only in a small part of the driving
period. It has been found that there is an $f$ range ($\sim 1-100$
mHz) where both EC and FD patterns can exist in each driving half
period in the form of successive (time shifted) flashes.
Theoretical calculations based on the standard model of EC
extended with flexoelectricity \cite{Krekhov2008} (which is able
to describe FDs too \cite{Krekhov2011}) have justified that
flashing patterns are indeed the solutions of the
nemato-electrohydrodynamic equations at ultralow $f$. The
calculated position of the FD flashes within the driving half
period showed quantitative matches with the experiments, while for
the position of the EC flashes the frequency dependence was only
qualitatively reproduced by the calculations, as the EC flashes
come earlier within the period than expected \cite{Eber2012}.

In this paper we present further experimental results on the
ultralow $f$ behaviour, however, in a different system than those
reported before. The paper is organized as follows.
Section~\ref{sec:exp} introduces our compound and the experimental
method. The new findings are grouped around three subtopics:
Section~\ref{sec:results:1} reports on the temporal evolution of
the patterns within the period; Section~\ref{sec:results:2} deals
with the frequency dependence of the threshold characteristics and
Section~\ref{sec:results:tempdep} provides data on the temperature
dependence of various material parameters. Finally the paper is
concluded in Section~\ref{sec:concl} with a summary and some
closing remarks.

\section{\label{sec:exp}Experimental}

\begin{figure}[h]
\centering
  \includegraphics[width=6cm]{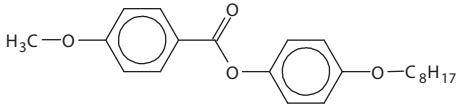}
  \caption{The chemical structure of the rod-like nematic molecule
  4-n-octyloxyphenyl 4-n-methyloxybenzoate (1OO8).}
  \label{fig:1oo8}
\end{figure}

Our measurements have been performed on the nematic liquid crystal
4-n-octyloxyphenyl 4-n-methyloxybenzoate (1OO8)\footnote{\textit{Two abbreviation styles are known in
the literature for the members of the 4-n-alkyloxyphenyl
4-n-alkyloxybenzoate homologous series. Here we have adopted the
one used by Nair et al. \cite{Nair2008} According to the
alternative style by Kochowska et al.\cite{Kochowska2004} the same
compound could also be abbreviated as 1/8.}} that
shows only a nematic mesophase. The chemical structure of 1OO8 is
shown in Fig.~\ref{fig:1oo8}.

In heating it melts to nematic from  the crystalline phase at 63.5
$^\circ$C, while the clearing point ($T_{NI}$) equals to 76.7
$^\circ$C. The nematic phase can be supercooled down to
53$^\circ$C. The material parameters of 1OO8, such as the
dielectric anisotropy
($\varepsilon_a=\varepsilon_{\parallel}-\varepsilon_{\perp}$), the
optical anisotropy ($n_a = n_{\parallel}-n_{\perp}$), the
anisotropy of the diamagnetic susceptibility
($\chi_a=\chi_{\parallel}-\chi_{\perp}$), and the bulk elastic
constants ($K_{11}, K_{22}, K_{33}$) were determined as the
function of temperature using a method based on magnetic and
electric Freedericksz-transitions \cite{Majumdar2011}. Here
$\varepsilon$ and $n$ denote the dielectric permittivity and the
refractive index, respectively; the subscripts $\parallel$ and
$\perp$ correspond to measurement directions parallel with and
perpendicular to the director.

The compound was investigated  in commercial sandwich cells
(E.H.C. Co.) with ITO electrodes coated with rubbed polyimide
layers for planar alignment. The electrode area was 1 cm$^2$. The
thickness of the empty cells ($d = 10.4 - 10.8$ $\mu$m) was
measured by an Ocean Optics spectrophotometer. During the
measurements the temperature of the sample was kept constant
within 0.01 $^\circ$C in an Instec HSi heat stage controlled with
an mK-1 board. The sample was driven by a sinusoidal voltage
$\tilde{U}(t)$ of an Agilent 33120A function generator via a
high-voltage amplifier: $\tilde{U}(t) = \sqrt{2} U \sin(2 \pi f t)$.

The electric field induced patterns were observed by a Leica DM RX
polarizing microscope in transmission mode with white light
illumination using the shadowgraph technique \cite{Rasenat1989}
(the polarizer was removed, while the analyser was set to be
parallel with the rubbing direction). The imaging system was
equipped with an EoSens MC1362 high speed camera interfaced by an
Inspecta-5 frame grabber. After waiting one or two periods of the
driving signal following the application of the voltage to the
sample (or waiting 5 seconds at frequencies higher than 0.2 Hz), a
sequence of 1000 images was recorded. The acquisition of the first
image was triggered by the zero crossing (from negative to positive) of the applied voltage.

In addition to the optical observations the electric current
through the cell was monitored by measuring the voltage drop on a
relatively small, known resistance connected in series with the
sample. Simultaneously the driving waveform was also recorded by a
TiePie Handyscope HS3 oscilloscope. The data acquisition and
processing system was fully automated.

\section{\label{sec:results}Results and discussion}

\subsection{\label{sec:results:1}Flashing contrast and current}

\begin{figure}[h]
\centering
\def\imagetop#1{\vtop{\null\hbox{#1}}}
(a)
\begin{tabular}{cc}
  \imagetop{\includegraphics[width=3.2cm]{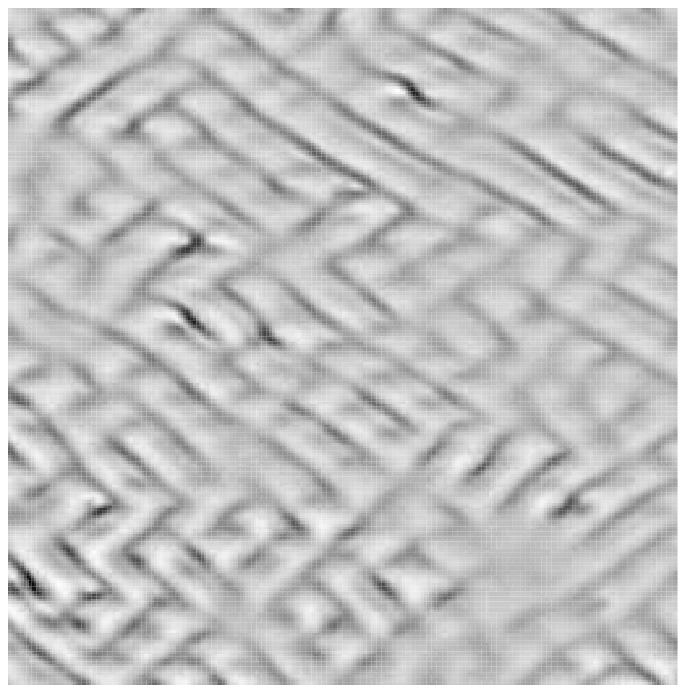}}&
  \imagetop{\includegraphics[width=3.8cm]{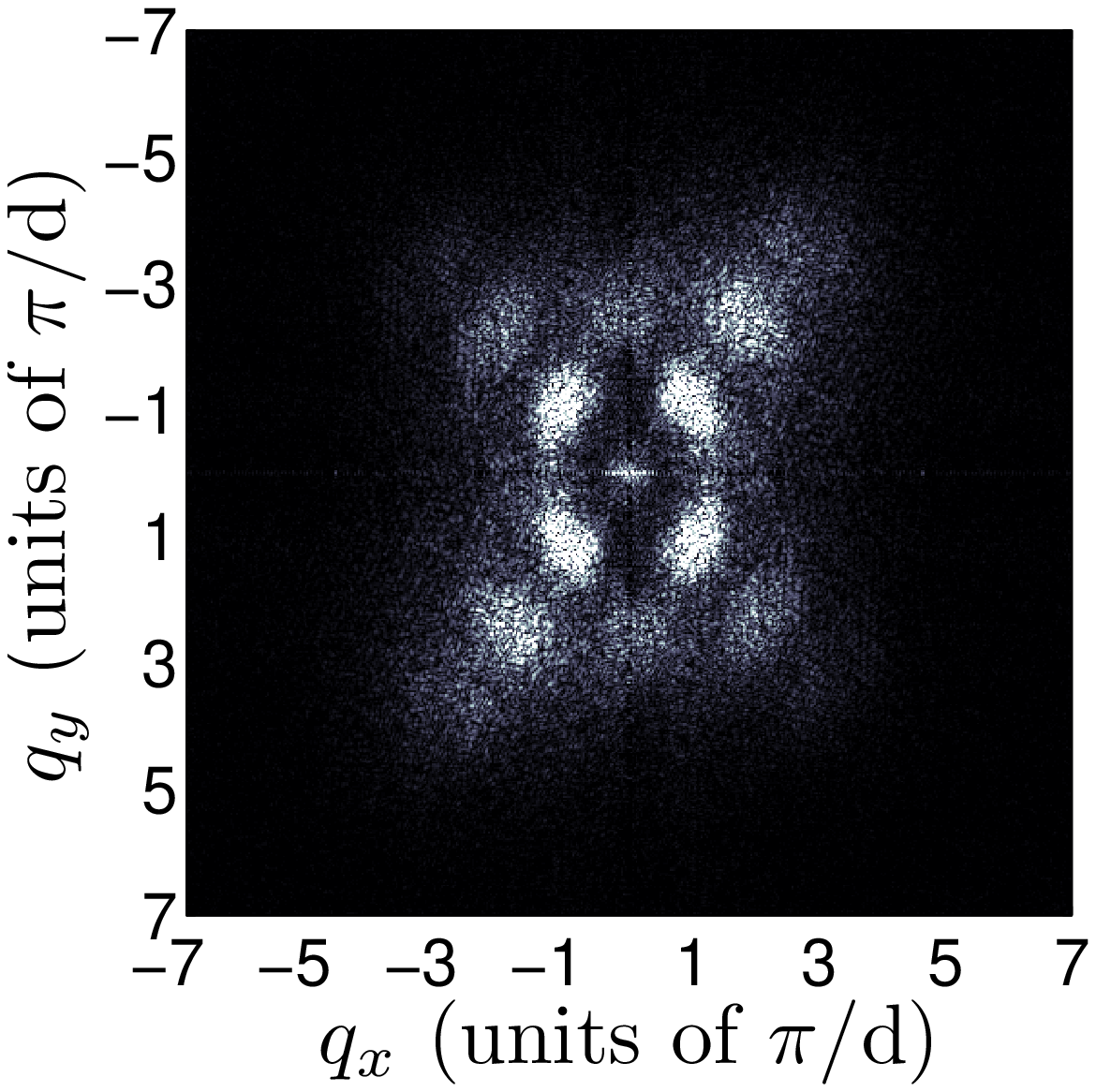}}
\end{tabular}
\\(b)
\begin{tabular}{cc}
  \imagetop{\includegraphics[width=3.2cm]{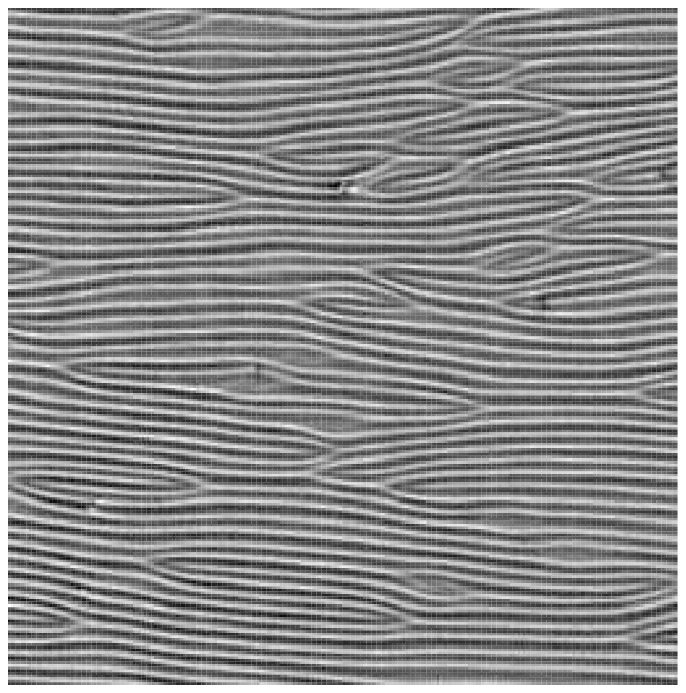}}&
  \imagetop{\includegraphics[width=3.8cm]{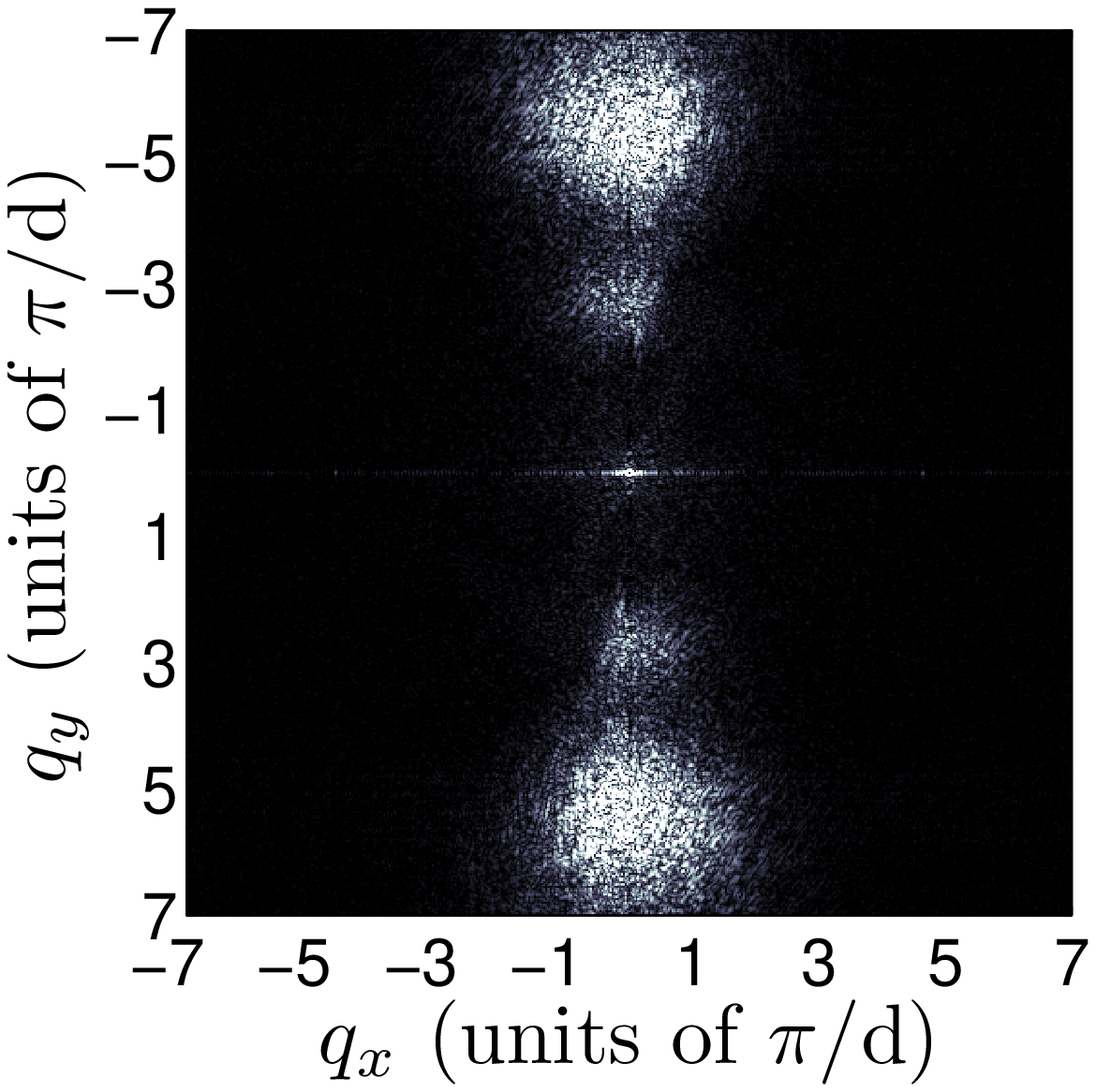}}
\end{tabular}
  \caption{Snapshot images and their 2-d Fourier transforms
  (a) for electroconvection and (b) for flexodomains at $f=50$ mHz and $U = 19$ V. The images cover 200 $\mu$m x 200 $\mu$m area. The initial director orientation lies horizontally.}
  \label{fig:patterns}
\end{figure}

Applying a low frequency (e.g. $f=50$ mHz)  sinusoidal voltage to
the cell, patterns appear above a threshold voltage in a narrow
time window in each half period of driving. Two distinct pattern
morphologies were found with different thresholds, similarly to
previous observations on other nematics \cite{Eber2012}.
Representative snapshots of the patterns and their 2-dimensional
(2-d) Fourier transforms (the spectral intensities) are presented
in Fig.~\ref{fig:patterns}. The two morphologies can be attributed
to oblique conductive EC rolls (a zig-zag pattern,
Fig.~\ref{fig:patterns}a) and to flexodomains
(Fig.~\ref{fig:patterns}b); the latter appear as stripes parallel
to the initial director alignment.

For a quantitative analysis of the pattern evolution it is
necessary to provide a proper definition for the pattern contrast,
which has a minimum (ideally zero) in the homogeneous state and
increases as the pattern emerges.

A common procedure is to perform a 2-d Fourier transformation of
the images in order to find the critical wave vector
$\mathbf{q_c}=(q_x, q_y)$ of the pattern (where the Fourier
amplitudes have maxima) and to define the contrast $C_q$ as the
sum of the spectral intensities in a region around $\mathbf{q_c}$.
It is clear from Fig.~\ref{fig:patterns} that the two pattern
types observed in 1OO8 (EC and FD) are characterized by different
$\mathbf{q_c}$ vectors, i.e. they are well separated in the
Fourier space. Therefore this contrast definition allows
distinguishing them not only from the initial homogeneous state,
but also between each other.

Alternatively, a mean square deviation of the image intensities,
$C_s=\langle (\Phi- \langle \Phi \rangle)^2 \rangle$, may also
serve as a measure of the contrast. Here $\Phi$ is the intensity
of an individual pixel and $\langle\rangle$ denotes averaging over
the whole image. This definition is simpler, though it has the
disadvantage of not being able to distinguish various pattern
morphologies. Actually $C_s$ would coincide with $C_q$ if the
summation of the spectral intensities were extended over the whole
Fourier space.

\begin{figure}[h]
\centering
  \includegraphics[width=8cm]{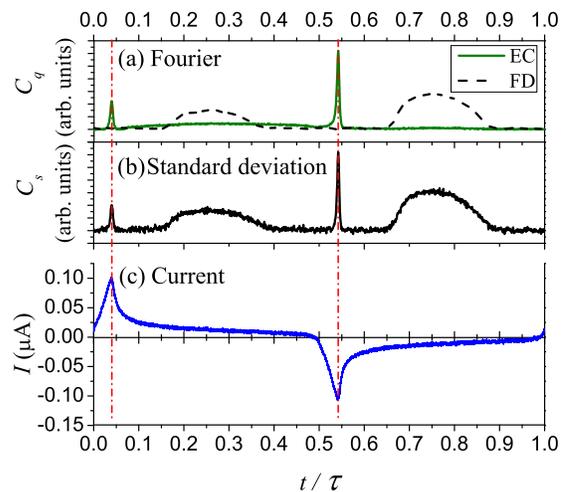}
  \caption{(Color online) The time dependence within a driving period
  (a) for the contrast $C_q$ obtained by Fourier technique;
  (b) for the contrast $C_s$ calculated from the square deviation; and
  (c) for the electrical current $I$ through the liquid crystal.
  $t=0$ corresponds to the zero crossing (from positive to negative) of the applied voltage. The dashed-dotted lines show that the peaks of EC and of the current coincide.}
  \label{fig:c_i_t}
\end{figure}

Figure~\ref{fig:c_i_t} exhibits and compares the time dependence
of contrast within a driving period for both definitions given
above, measured in a $d = 10.4\;\mu$m thick cell at $T-T_{NI} =
21.7$ $^\circ$C driven by an $f = 22$ mHz, $U = 18$ V
voltage. Figure~\ref{fig:c_i_t}a shows $C_q$ obtained by the
Fourier method for the EC (solid line) and the FD (dashed line)
patterns. Both curves exhibit a single peak in each half period,
but at different time intervals; hence these two pattern types are
well separated not only in the Fourier space, but in time as well.
In Fig.~\ref{fig:c_i_t}b the contrast $C_s$ calculated by the
square deviation is plotted. This curve has two, well separated
peaks per half period (looks similar to the superposition of the
two curves in Fig.~\ref{fig:c_i_t}a); thus can also be used to
detect the appearance of both pattern types. Therefore, for
simplicity, in the following we will use $C_s$ as the measure of
the contrast of the patterns.

Figure~\ref{fig:c_i_t}c depicts the time dependence of the
electrical current which was measured simultaneously with image
acquisition. At this $f$ and $U$ the current is highly
nonlinear; it can be characterized by sharp peaks rather than by a
harmonic response. It can be deduced from the figure that,
surprisingly, the location of the maxima of the current peaks
coincide precisely with the contrast peaks corresponding to the EC
flashes (see the dash-dotted vertical lines in
Fig.~\ref{fig:c_i_t}). Numerous different voltages, frequencies
and temperatures were tested. Though at various conditions the
time instant of the EC flash may change \cite{Eber2012}, it still
equals to that of the current peak; thus we can conclude that this
is not an accidental coincidence. We suggest that the current
spikes trigger the emergence of the EC pattern. Therefore it
appears earlier within the half period (a phase-locking behaviour)
than expected otherwise.

We note that the spiky behaviour of the current is not a
consequence of the appearance of the EC pattern. Current spikes
have been detected at low voltages (much below any pattern
threshold) where no patterns are observable and also in the
isotropic phase. We think that the nonlinear current behaviour is
due to ionic effects and to the presence of insulating polyimide
orienting layers on the electrode surfaces of the cell. The
presence of (relatively low) concentration of ionic impurities in
the nematic makes it to behave as a weak electrolyte. In the
studied ultralow frequency range the current due to the linear
impedance of the cell (i.e. the capacitive and the ohmic
components) is at least an order of magnitude smaller than the
transient currents due to building or destroying the Debye
screening layers near the electrodes; the latter occurs at each
polarity reversal of the voltage.

In order to describe the behaviour of weak electrolytes in
electric fields several models were developed, differing in their
sets of assumptions
\cite{Derfel2009,Freire2006,Freire20062,AlexeIonescu2009,Palomares2004,Barbero2006,Barbero20062,Barbero2008,Atasiei2008};
i.e. they take into consideration different subsets of the
possible effects listed below: generation and recombination of
ions; different mobilities, diffusion coefficients and charges of
ionic species; surface adsorption; charge injection; chemical
reactions; voltage attenuation due to the orienting layers; etc.
Due to the complexity of the models they mostly focused on the
linear response and calculated the low frequency complex impedance
which could be compared to low $f$ dielectric spectroscopy data.

Recently theoretical calculations of the nonlinear current
characteristics in response to a low frequency sinusoidal voltage
driving were also reported \cite{Derfel2009,Freire2006}, yielding
curves similar to those shown in Fig.~\ref{fig:c_i_t}c, however,
without comparison with experiments. This gives the hope that
after measurements or intelligent guesses of the unknown material
parameters of the model the measured current response can be
reproduced; it is remaining a task for the future.

The nematic being a weak electrolyte has consequences on the
pattern formation processes. It was shown that the weak
electrolyte model (WEM) of EC \cite{Treiber1995}, which considers
ionic dissociation and recombination, can account for the
travelling of EC roll patterns found occasionally at frequencies
above a few tens Hz. This model has not yet been analysed for low
driving frequencies; due to its high complexity it remains a
challenge for the future to decide whether it is able to describe
the phase locking of EC flashes to current spikes.

\subsection{\label{sec:results:2}Threshold characteristics}

Flexodomains and electroconvection both are threshold phenomena;
i.e. the patterns with a critical wavenumber
$q_{c}=|\mathbf{q_c}|$ occur above a threshold voltage $U_{c}$.
Determination of $U_{c}$ and $q_{c}$ is therefore the primary task
at pattern characterization. At high frequencies ($f>10$ Hz) for
$U>U_{c}$ patterns usually develop within seconds; therefore
thresholds can easily be estimated by increasing $U$ as the
voltage at which the pattern becomes perceptible by eyes in the
microscope. This simple technique practically does not work at our
ultralow frequency driving, since the driving period is quite long
and in addition the patterns appear as flashes, which means they
can be observed only in a short time window.

In order to determine $U_{c}$ precisely one has to follow
quantitatively the emergence of patterns from the homogenous
state; i.e. to record and then analyse the contrast-voltage
curves. As the contrast varies within the driving period (as shown
in Fig.~\ref{fig:c_i_t}b), the maximum $C_m$ of the contrast $C_s$
in the FD (or EC) peak can be regarded as a measure to what extent
the FD (or EC) pattern has been developed at a given applied
voltage.

In an ideal case (perfect bifurcation) the contrast $C_m$ should
be zero at voltages below the threshold. Experimentally a nonzero
background contrast $C_b$ is always found even in the homogeneous
state at no applied voltage ($C_b=C_s(U=0)$). This background
contrast comes from various sources: the electronic noise of the
camera, the thermal fluctuation of the director in a planar
nematic, imperfections of the orientation or inhomogeneity of the
illumination. This background was automatically subtracted from
each data point; thus it will not be indicated in the forthcoming
figures.

As the voltage is increased above $U_c$, the initial planar
director orientation ${\bf n}_0 = (1,0,0)$ becomes unstable and a
spatially periodic director distortion $\delta {\bf n} = {\bf
n}_{lin} A \exp[i(q_x x + q_y y)]$ appears. Here ${\bf n}_{lin} =
(0,n_y,n_z)$ is a linear eigenvector, $A \propto \sqrt{U^2 -
U_c^2}$ characterize the amplitude of the distortion, and ${\bf
q_c} = (q_x,q_y)$ is the wavevector of the pattern. The spatially
periodic director distortion results in a shadowgraph image whose
intensity modulation $I_s$ depends on the amplitude of the
vertical distortion $A n_z$. For small distortion amplitudes (not
too far from threshold) the intensity modulation in the leading
order is given\cite{Trainoff:2002} by $I_s = c_a A + c_p A^2$ with
the first order amplitude term and the second order phase term.
For EC patterns (normal rolls with ${\bf q_c} = (q_x,0)$) the
linear term is dominating and $I_s \propto A$. In case of FD,
where ${\bf q_c} = (0,q_y)$, the relevant contribution to the
shadowgraph intensity is of the second order: \cite{Pesch:private}
$I_s \propto A^2$. The contrast of the shadowgraph image defined
as the mean square deviation of the image intensities is then $C_s
\propto I_s^2$. Thus the maximum of the contrast within the
driving period is expected to be $C_{mEC} \propto (U^2 -
U_{cEC}^2)$ for an EC pattern and $C_{mFD} \propto (U^2 -
U_{cFD}^2)^2$ for the FD \cite{May2008}. In the vicinity of the
threshold $(U^2 - U_{cFD}^2) \approx 2 U_c (U-U_c)$; therefore
$C_{mEC}$ as well as $\sqrt{C_{mFD}}$ should grow linearly with
the voltage.

\begin{figure}[h]
\centering
  \includegraphics[width=8cm]{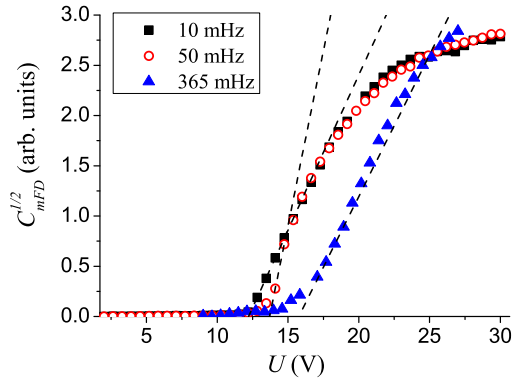}
  \caption{(Color online) The voltage (rms) dependence of the square root of the FD
  contrast peaks for different frequencies (symbols). The dashed lines
  indicate the linear extrapolation.}
    \label{fig:c_u_fd}
\end{figure}

Figure~\ref{fig:c_u_fd} shows the measured $\sqrt{C_{mFD}}(U)$
curves for a few frequencies. It is seen that the linear relation
near the threshold is obeyed quite well; though the transition is
smeared a little (due to imperfections and/or the occurrence of
subcritical fluctuations). Therefore the threshold voltage
$U_{cFD}$ is actually determined by a linear extrapolation, as the
intersection of the horizontal axis with the line fitted onto the
linear section of the $C_m(U)$ curve slightly above the suspected
threshold. This procedure is going to be referred as method A.

\begin{figure}[h]
\centering
  \includegraphics[width=8cm]{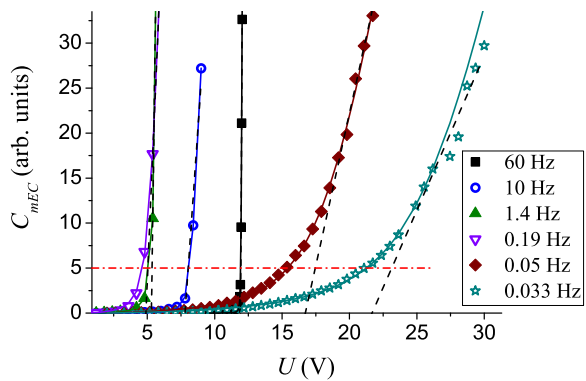}
  \caption{(Color online) The voltage (rms) dependence of the contrast peaks $C_m$ of EC for different
   frequencies (symbols). Solid lines are fits with the imperfect bifurcation model,
   the dashed lines indicate the linear extrapolation.}
    \label{fig:c_u_ec}
\end{figure}

The voltage dependence of $C_{mEC}$ for EC is shown in
Fig.~\ref{fig:c_u_ec} for several driving frequencies. It is
clearly seen that the frequency affects not only the threshold
voltages, but also the character (the shape) of the $C_{mEC}(U)$
curves. Evidently the linear relation holds only at high
frequencies; there the thresholds $U_{cECA}$ can be determined by
extrapolation (method A).

Below 1 Hz, however, there is no sharp increase of the contrast;
the $C_m(U)$ curves show rather a slow gradual increase, while the
contrast levels and thus the visibility of the patterns vary in
the same range as at high frequencies. The determination of
thresholds is then not so straightforward. In lack of a well
defined linear part of the contrast curve, method A becomes
unreliable; the choice of points used for the extrapolation (the
dashed lines in Fig.~\ref{fig:c_u_ec}) is to some extent
arbitrary.

An alternative way (method B) is to select (arbitrarily) a
critical contrast value $C_0$ (the dash-dotted line in
Fig.~\ref{fig:c_u_ec}) where the EC pattern is visible by eye. The
voltage $U_{cECB}$, where $C_{mEC}(U_{cECB})=C_0$, can be regarded
as another estimate of the threshold. In case of forward
bifurcations, which the standard EC pattern formation is an
example for, the contrast increases continuously from zero.
Therefore $U_{cECB}$ slightly overestimates the threshold.

The change in the shape of the $C_{mEC}(U)$ curves may be
interpreted so that the nearly perfect bifurcation (at high $f$)
becomes imperfect at lower $f$. For an imperfect bifurcation the
amplitude of the director distortion $A$ satisfies the equation

\begin{equation}
\label{eq:ampleq1} \varepsilon A - g A^3 + \delta = 0.
\end{equation}

Here $\varepsilon = U^2/U_{cEC}^2-1$, $U$ is the rms applied
voltage, $U_{cEC}$ is the threshold voltage, $g>0$ characterizes
the saturation of the amplitude and $\delta \geq 0$ is the measure
for the imperfection ($ \delta = 0$ corresponds to the perfect
forward bifurcation). For $g > 0$ and $\delta > 0$ only one of the
three solutions of Eq.~(\ref{eq:ampleq1}) is stable in the whole range of $\varepsilon > -1$ and thus
relevant; it reads as

\begin{equation}
\label{eq:ampleq2}
\begin{split}
A = (\frac{\delta}{2g})^{1/3} F(\tilde{\varepsilon}),\\
F(\tilde{\varepsilon})=(\frac{\tilde{\varepsilon}}{\hat{f}(\tilde{\varepsilon})}
+\hat{f}(\tilde{\varepsilon})) \; \text{for} \; \tilde{\varepsilon} \leq 1,\\
F(\tilde{\varepsilon}) = 2 \sqrt{\tilde{\varepsilon}}
\cos{ ( \frac{1}{3}\arctan{(\sqrt{\tilde{\varepsilon}^3-1})} ) }
\; \text{for} \; \tilde{\varepsilon}>1,\\
\tilde{\varepsilon} = \frac{2}{3} \frac{\varepsilon}{(2 g \delta^2)^{1/3}},
\; \hat{f}(\tilde{\varepsilon}) = (1+\sqrt{1-\tilde{\varepsilon}^3})^{1/3}.
\end{split}
\end{equation}

\begin{figure}[h]
\centering
  \includegraphics[width=7cm]{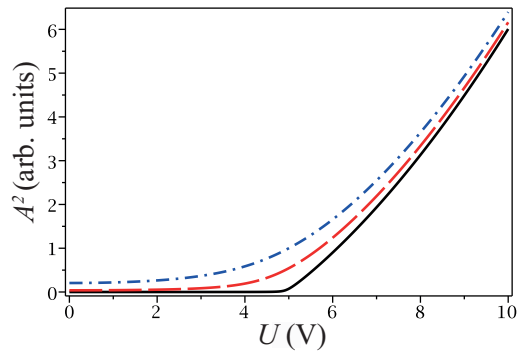}
  \caption{(Color online) Square of the pattern amplitude $A^2$ as a function
  of the applied voltage $U$ for $\delta = 0.01$ (solid line),
  $\delta = 0.2$ (dashed line), and $\delta = 0.5$ (dot-dashed line).
  $U_{cEC} = 5$, $g = 0.5$.}
    \label{fig:a02_u}
\end{figure}

As mentioned above, the maximum contrast $C_{mEC}$ of the EC
patterns observed using the shadowgraph technique is proportional
to $A^2$. In Fig.~\ref{fig:a02_u} the dependence of $A^2$ on the
applied voltage $U$ is shown for different values of the
imperfection parameter $\delta$ at fixed values of $U_{cEC}$ and
$g$. It demonstrates that the shape of the curve changes
substantially if the imperfection ($\delta$) increases.

For a precise quantitative analysis we can use the same background
subtraction here, just as was done with the experimental data;
therefore the contrast depicted in Fig.~\ref{fig:c_u_ec} will be
related to the amplitude as:

\begin{equation}
\label{eq:impfit} C_{mEC}=C_{max}-C_b = \alpha [A^2(U)-A^2(U=0)],
\end{equation}
where $C_{max}$ is the maximum contrast of the pattern, $C_b$ is
the background contrast at $U=0$, and $\alpha>0$ is a scaling
factor. Combining Eqs.~(\ref{eq:ampleq2}) and (\ref{eq:impfit})
one can fit the experimental $C_{mEC}(U)$ curves by this
phenomenological model for imperfect bifurcation using four
parameters: $\alpha$, $g$, $\delta$ and $U_{cEC}$ (method C).

The actual value of the scaling parameter $\alpha$ is determined
by the optical set-up and the optical properties. As $q_c$ of the
EC pattern depends weakly on $f$, we can assume that $\alpha$ is
frequency independent. Its value could be obtained from the fit at
$f=60$ Hz, leaving only three free parameters for the fits at
lower frequencies.

\begin{figure}[h]
\centering
  \includegraphics[width=8cm]{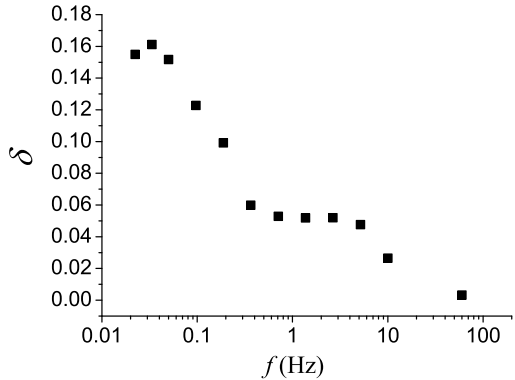}
  \caption{Frequency dependence of the imperfection parameter $\delta$.}
  \label{fig:delta_f}
\end{figure}

The results of the fit procedure are shown by solid lines in
Fig.~\ref{fig:c_u_ec}. The match with the experimental data are
quite convincing. The frequency
dependence of the imperfection parameter $\delta$ is plotted in
Fig.~\ref{fig:delta_f}. It clearly shows -- what we have already
expected from the experimental data in Fig.~\ref{fig:c_u_ec} --
that the imperfection grows at lower frequencies. Several reasons
could be responsible for the increase of the apparent
imperfection.

In planar samples aligned by rubbed polyimide layers a small
director pretilt at the confining plates is practically
unavoidable. Such pretilt is known to yield imperfect bifurcation
(i.e. lack  of a sharp threshold) in the case of splay
Freedericksz-transition. The effect of a tilted alignment on the
EC characteristics has theoretically been studied only for high
frequencies \cite{Hertrich1995}; the pretilt modified $U_c$, but
did not affect the sharpness of the threshold, which is in
agreement with our observations (Fig.~\ref{fig:c_u_ec}) at high
$f$.

Decreasing the frequency of the applied ac voltage well below the
inverse director relaxation time may, however, alter the situation
as one enters the regime of quasistatic director response. Here a
small pretilt may enhance the director deformations and
correspondingly the contrast of the pattern can develop already at
lower voltage amplitudes compared to the high frequency case.
Unfortunately a detailed theoretical analysis of this regime in
the presence of pretilt is not yet available.

The nonlinear electric current characteristics presented in
Sec.~\ref{sec:results:1} may provide another reason for the
apparent softening of the ultra-low $f$ EC thresholds. The
coincidence of the electric current peaks and the EC flashes
clearly shows the strong correlation between pattern formation and
ionic phenomena: the massive ionic flow helps the
electro-hydrodynamical instability to emerge. The spatial
distribution of the current is not necessarily uniform, mainly due
to surface inhomogeneities (which may originate e. g. from
crystallization of the compound) or small variations in the cell
thickness and/or pretilt. The current inhomogeneities may locally
reduce the threshold of EC. In fact this effect has been observed:
the EC pattern first appears in germs and extends gradually to
larger area by increasing the voltage. The location of these germs
can be identified even in the well developed pattern as small
spots/patches of higher contrast (a few such spots can be seen in
Fig.~\ref{fig:patterns}a). The contrast $C_{mEC}(U)$ of the
pattern plotted in Fig.~\ref{fig:c_u_ec} is calculated over the
whole image; thus a continuous increase of the area filled with
pattern leads to a continuous increase of $C_{mEC}(U)$.
Consequently a locally sharp transition yields a softened, gradual
contrast variation. While ionic effects are mostly negligible at
high frequencies (linear current response), they become crucial at
ultralow frequencies (spiky current response), which may explain
the increase of the imperfection parameter for $f \rightarrow 0$.

We note that the formation of flexodomains is not affected by the
electric current spikes as they occur in different time windows.
Therefore the above scenario of germ-induced pattern evolution
does not apply to FD; i.e. the onset of FD remains sharp over the
full frequency range of its existence, as shown in
Fig.~\ref{fig:flexo}.

\begin{figure}[h]
\centering
  \includegraphics[width=8cm]{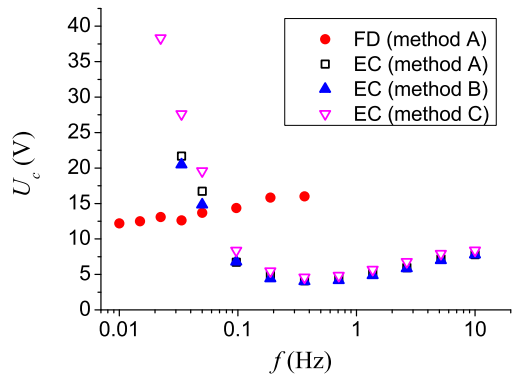}
  \caption{(Color online) The frequency dependence of threshold voltages (rms) of EC
  and FD determined by various methods.}
  \label{fig:uth_f}
\end{figure}

The frequency dependence of the threshold voltages of both
patterns can be seen in Fig.~\ref{fig:uth_f}. It depicts the
$U_{cEC}$ values determined by all three methods introduced above.
The data by methods A (extrapolation) and B (comparison) almost
coincide, while the thresholds obtained from fitting to the
imperfect bifurcation model are significantly larger at lower
frequencies. This is not surprising since methods A and B
intrinsically assume that no deformation exists below a threshold,
while an imperfect bifurcation actually means a thresholdless
deformation with $U_c$ being a parameter only.

Otherwise the $U_{cEC}(f)$ curve exhibits the expected behaviour.
The reduction of the threshold at lowering $f$ in the $0.5<f<10$
Hz range corresponds to the theoretical predictions and matches
the behaviour of other nematics \cite{TothKatona2008}. The
increase of $U_{cEC}$ toward ultralow frequencies is attributed to
the internal attenuation due to the insulating polyimide alignment
layers on the electrodes \cite{Eber2012}. The frequency dependence
of $U_{cFD}$ seems to be significantly weaker than that of
$U_{cEC}$ in the same $f$ range. Taking into account the internal
attenuation, the actual FD threshold voltage (on the liquid
crystal layer) grows much stronger with $f$ than the apparent
threshold plotted in the figure (the voltage applied to the cell),
which is in agreement with the theoretical
predictions \cite{Krekhov2011}.

Figure~\ref{fig:uth_f} clearly shows that the two distinct
patterns, EC and FDs, coexist in a relatively wide (0.02 Hz $<f<$
0.4 Hz) range, even though their threshold voltages are quite
different. This is possible, because they remain separated in time
until the half period of driving voltage is large enough for both
patterns to emerge and decay; thus they can build up from the same
almost homogenous initial state. For $f>0.4$ Hz, however, this
does not hold any more. In that range, besides the shorter period
time, $U_{cEC}$ is much lower than $U_{cFD}$. Thus the EC contrast
spikes become much broader and the EC pattern does not decay fully
before FD should emerge. Under such condition the FD pattern
(which has a lower contrast than EC) cannot be recognized any
more.

As the frequency is reduced, at around $0.05-0.07$ Hz there is an
intersection of the two threshold curves ($U_{cFD}$ and
$U_{cEC}$). At $f$ below this intersection the threshold of FDs is
lower than that of EC; thus upon increasing the voltage FD is the
first instability, EC sets on at a higher voltage. This is in
accordance with the finding that when applying a pure DC voltage,
no EC pattern, only FDs can be detected.

\begin{figure}[h]
\centering
  \includegraphics[width=8cm]{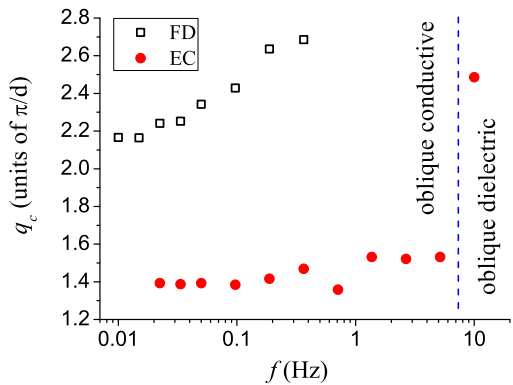}
  \caption{(Color online) The frequency dependence of the threshold wave numbers for EC and FD.}
  \label{fig:qth_f}
\end{figure}

Characterization of the threshold behaviour is incomplete without
addressing the frequency dependence of the critical wave number
$q_c=|\mathbf{q_c}|$. Figure~\ref{fig:qth_f} exhibits the relevant
curves both for EC ($q_{EC}$) and FD ($q_{FD}$). The values were
determined using the 2-d fast Fourier transformation (FFT) of
images taken slightly above the threshold, at $U=1.05 U_c$, in
order to have sufficient contrast for the evaluation. Note that
for the oblique EC rolls $q_{EC}=\sqrt{q_x^2+q_y^2}$, while FDs
are parallel to the initial director, so $q_{FD}\approx q_y$. The
wave numbers increase for both patterns with the frequency. In the
case of FD there is a moderate $f$ dependence even at ultralow
frequencies. For EC, the change of $q_{EC}$ seems to be very small
until 5 Hz. Between 5 and 10 Hz, however, the wave number
increases suddenly, which is attributed to the transition between
oblique conductive and oblique dielectric EC. To our knowledge no
such transition was reported before in the literature. We note
that the obliqueness angle decreases with the frequency, the
Lifshitz-point is reached in the dielectric regime at $f_{L}
\approx 80$ Hz.

\subsection{\label{sec:results:tempdep}Temperature dependence of the flexoelectric coefficients}

Though several experimental methods have been proposed to measure
the flexoelectric coefficients, measurements usually cannot be
done without serious compromises \cite{Buka2012}. Analysis of the
threshold parameters ($q_{cFD}$, $U_{cFD}$) of the flexoelectric
instability is one of the possible methods. Its drawback is that
only a few compounds exhibit this effect, because: 1) the material
needs to have a quite low dielectric anisotropy
($|\varepsilon_a|\ll 1$), 2) the concentration of its ionic
impurities should be sufficiently low in order to avoid large
screening effects, 3) other phenomena (e.g. EC or
Freedericksz-transition) should not influence the homogenous
planar initial state below the threshold of FD.

The threshold characteristics for dc driving voltage have long ago
been calculated analytically \cite{Bobylev1977} using the
one-elastic-constant approximation ($K_{11}=K_{22}=K$):

\begin{equation}
\label{eq:pikinu} \tilde{U}_{cFD} = \frac{2 \pi K}{|e_{1}-e_{3}|
(1+\mu)},
\end{equation}

\begin{equation}
\label{eq:pikinq} \tilde{q}_{cFD} = \frac{\pi}{d}
\sqrt{\frac{1-\mu}{1+\mu}},
\end{equation}
where $e_{1}$ and $e_{3}$ are the splay and bend flexoelectric
coefficients, respectively, and
\begin{equation} \label{eq:mu}
\mu = (\varepsilon_0 \varepsilon_a K)/|e_{1}-e_{3}|^2.
\end{equation}

According to Eq.~(\ref{eq:pikinq}) the flexodomains can only exist
for the material parameter combination $|\mu| < 1$. This leads to
the requirement $|\varepsilon_a|<|e_{1}-e_{3}|^2/(\varepsilon_0
K)$ that should be valid for materials showing FDs. Combining
Eqs.~(\ref{eq:pikinq}) and (\ref{eq:mu}) yields:

\begin{equation}
\label{eq:estar} |e_{1}-e_{3}|=\sqrt{\varepsilon_0 \varepsilon_a K
\frac{1+q_{cFD}^2}{1-q_{cFD}^2}}.
\end{equation}

\begin{figure}[h]
\centering
  \includegraphics[width=8cm]{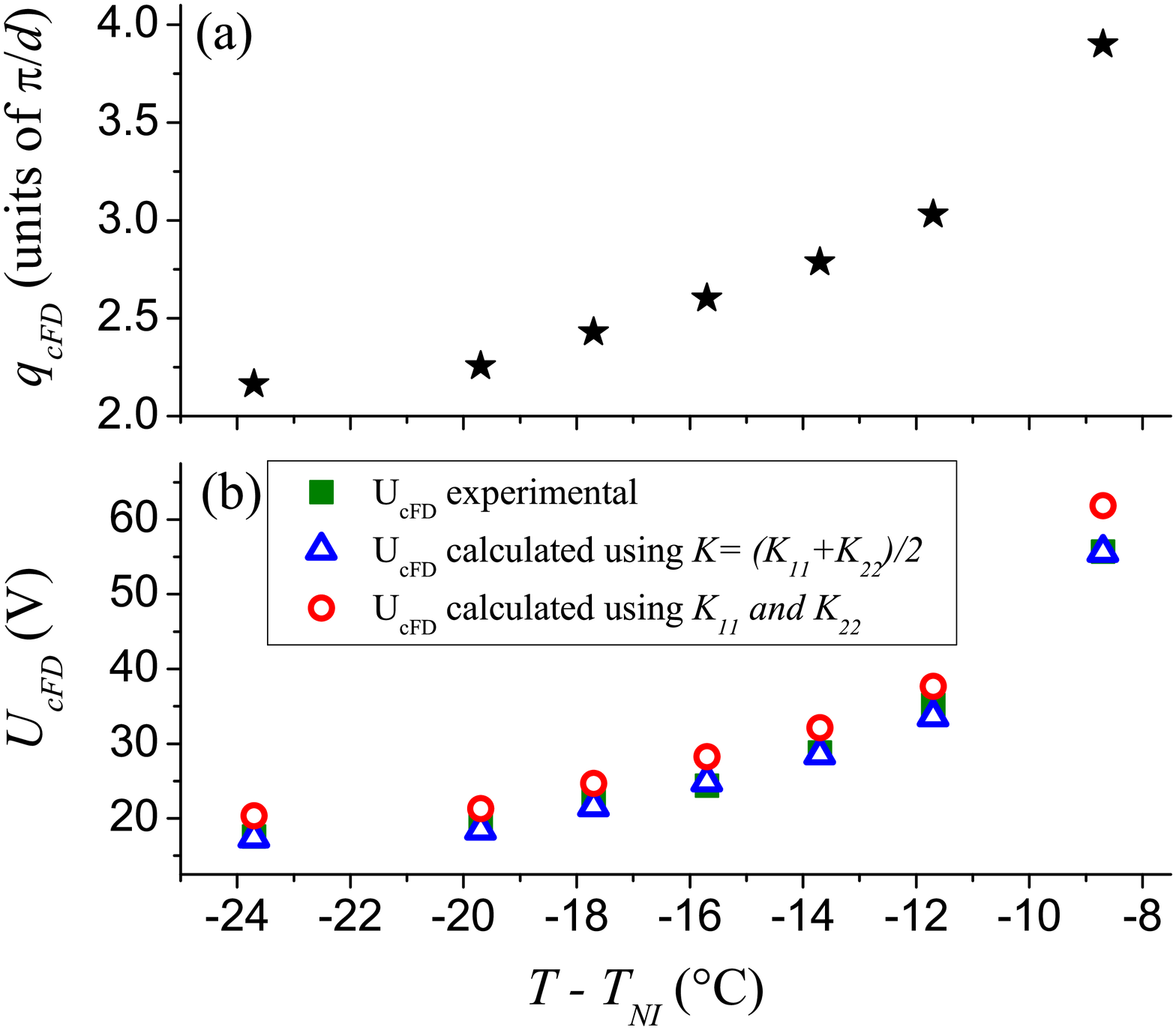}
  \caption{(Color online) The temperature dependence of (a) the wave number $q_{cFD}$
  and (b) the voltage $U_{cFD}$ (amplitude) at the onset of flexodomains.}
  \label{fig:q_u_temp}
\end{figure}

For 1OO8 both $q_{c FD}$ and $U_{c FD}$ were measured as the
function of temperature using 10 mHz ac sine voltage. We assumed that 10 mHz is low frequency enough to be considered as a quasistatic case, hence we have fitted the results with a static model. Therefore $U_{c FD}$ here is presented in voltage amplitude values instead of rms, since FD appears when the driving voltage reaches its maxima. Therefore $U_{cFD}$ in
Fig.~\ref{fig:q_u_temp} is presented in voltage amplitudes instead
of rms values.

Both $q_{cFD}$ and $U_{cFD}$ increase strongly toward higher
temperatures. Above $T-T_{NI} = -8$ $^\circ$C we could not detect
flexodomains up to the voltage of 135 V.

\begin{figure}[h]
\centering
  \includegraphics[width=8cm]{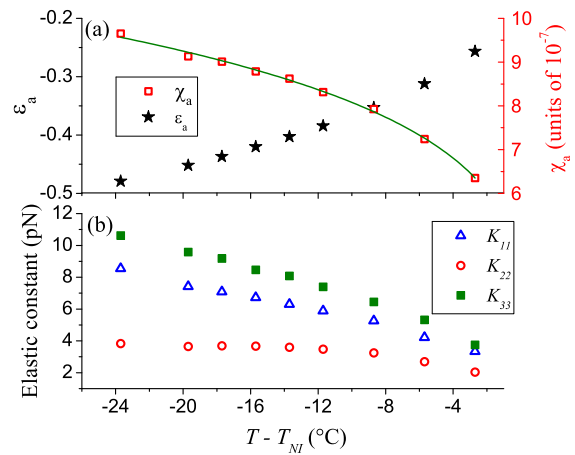}
  \caption{(Color online) The temperature dependence of (a) the dielectric ($\varepsilon_a$)
  and the diamagnetic ($\chi_a$) anisotropies, (b) the three elastic moduli.}
  \label{fig:tempdep}
\end{figure}

In order to determine $|e_{1}-e_{3}|$ we have measured some
material parameters of 1OO8 using methods based on electric and
magnetic Freedericksz-transitions. The temperature dependence of
$\varepsilon_a$ and of the diamagnetic susceptibility anisotropy
($\chi_a$) is shown in Fig.~\ref{fig:tempdep}a. $\varepsilon_a$ is
negative and relatively small, as it was expected. Therefore in
our planar sandwich cell geometry the dielectric interaction
stabilizes the planar structure; no electric field induced
Freedericksz-transition occurs. The values and the thermal
behaviour of $\chi_a$ are in the regular range of those in
rod-like nematics. This also holds for the elastic constants
$K_{11}$, $K_{22}$, and $K_{33}$, which are plotted in
Fig.~\ref{fig:tempdep}b. We note that $K_{33}$ is shown only for
the sake of completeness; we do not use it further on.

\begin{figure}[h]
\centering
  \includegraphics[width=8cm]{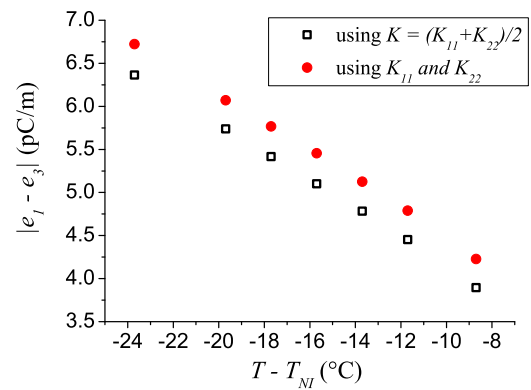}
  \caption{(Color online) The temperature dependence of the combination $|e_1-e_3|$
  of the flexoelectric coefficients.}
  \label{fig:flexo}
\end{figure}

The temperature dependence of $|e_{1}-e_{3}|$, presented in Fig.
\ref{fig:flexo}, was calculated from the measured data by two
different techniques. The first method (square symbols) was based
on the analytical formula, Eq.~(\ref{eq:estar}), of the
one-elastic-constant approximation, taking $K =
(K_{11}+K_{22})/2$. The second technique (triangle symbols)
utilized the recent theory \cite{Krekhov2011} of flexoelectric
domains that takes into account the anisotropic elasticity
($K_{11} \neq K_{22}$), calculating $|e_{1}-e_{3}|$ numerically.
As seen in Fig.~\ref{fig:flexo}, the second method provided values
about 7\% higher than those by the first one; both values of
$|e_{1}-e_{3}|$ fall in the regular range of that of rod-like
nematics.

In order to check the consistency of our models and the obtained
data, we have calculated $U_{cFD}$ using the $|e_{1}-e_{3}|$
values determined from $q_{cFD}$. The results, depicted in
Fig.~\ref{fig:q_u_temp}b, show that the first model gave about 2\%
lower, while the second one about 11\% higher values for
$U_{thFD}$ than the experiments.

Knowing the temperature dependence of $|e_{1}-e_{3}|$ gave us an
opportunity to compare our results with the predictions of the
molecular theory of flexoelectricity. It is expected
\cite{Helfrich1971,Derzhanski1971,Osipov2012} that the difference
of flexoelectric coefficients should be proportional to the square
of the order parameter $S(T)$:
\begin{equation}
\label{eq:es} |e_{1}-e_{3}|= \hat{e} S^2(T),
\end{equation}
where the proportionality constant is denoted by $\hat{e}$.

In Fig.~\ref{fig:flexo} $|e_{1}-e_{3}|$ is decreasing with the
temperature, which is consistent with the similar tendency of the
order parameter. For a more quantitative comparison, the knowledge
of $S(T)$ would be essential. $S(T)$ can only be accessed via
measuring physical quantities that are directly coupled to it. The
diamagnetic susceptibility, which is already determined from the
Freedericksz-transition measurements (Fig.~\ref{fig:tempdep}a) is
a good candidate, since it should be proportional to $S$ \cite{Stannarius1998}:
\begin{equation}
\label{eq:chias} \chi_a(T)= \hat{\chi} S(T),
\end{equation}
where $\hat{\chi}$ is a constant. In order to determine $\hat{\chi}$, and $S(T)$, the generalized form of the empirical Haller-extrapolation \cite{Leenhouts1979,Stannarius1998} method is applied, via fitting the experimental data of $\chi_a(T)$ with:
\begin{equation}
\label{eq:chiat} \chi_a(T)= \hat{\chi} \left(1-\beta \frac{T}{T_{NI}} \right)^\gamma,
\end{equation}
where $\beta$, $\gamma$ are constants, and the temperature data ($T$, $T_{NI}$) is measured in the Kelvin-scale. The result of the fit can be seen in Fig.~\ref{fig:tempdep}a (solid line). The parameters of the best fit correspond to: $\hat{\chi}=1.64 \times 10^{-6}$, $\beta=1$, and $\gamma =0.2$. Besides the dimensionless SI quantity of $\hat{\chi}$, its molar version is often used: $\hat{\chi}^M=\hat{\chi}M_m/\rho$, where $M_m$, and $\rho$ are the molar weight, and the density, respectively. Using $M_m=356.5~\mathrm{\frac{g}{mol}}$, and $\rho=1~\mathrm{\frac{g}{cm^3}}$ one gets $\hat{\chi}^M=585\times10^{-6}~\mathrm{\frac{cm^3}{mol}}$, that value fits well in the range of earlier results \cite{Stannarius1998,Ibrahim1979} obtained for different compounds with two aromatic rings.

Combining Eq.~(\ref{eq:es}) and Eq.~(\ref{eq:chias}) yields:
\begin{equation}
\label{eq:chiae} \chi_a = a \sqrt{|e_{1}-e_{3}|},
\end{equation}
with $a=\hat{\chi}/\sqrt{\hat{e}}$.

\begin{figure}[h]
\centering
  \includegraphics[width=8cm]{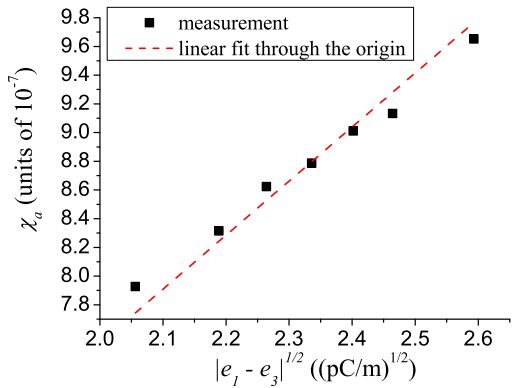}
  \caption{(Color online) The relation between $\chi_a$ and $\sqrt{|e_{1}-e_{3}|}$.}
  \label{fig:chia_sqrte}
\end{figure}

Figure~\ref{fig:chia_sqrte} provides a test of this relation, as
it plots the measured $\chi_a$ values against
$\sqrt{|e_{1}-e_{3}|}$ calculated for the same temperatures
(determined from the model with anisotropic elasticity). The fit
corresponding to Eq.~(\ref{eq:chiae}), represented by the dashed
line, seems to be quite good in spite of the fact that there was
only one fit parameter. The best fit results $a=0.38 (C/m)^{-0.5}$, that with $\hat{\chi}$ determined above, yields $\hat{e}=18.6$ pC/m.

\section{\label{sec:concl}Conclusions}

We have investigated the pattern forming phenomena induced by
ultralow frequency sinusoidal voltages applied onto the calamitic
nematic liquid crystal 1OO8. It was found that the behaviour in
this low frequency range is characteristically different from that
typical for high frequencies: here patterns appear as flashes in a
short time interval within each half period of driving. Two kinds
of pattern morphologies were detected: electroconvection rolls and
flexodomains. The types of patterns differ in their wave
vector (EC rolls are oblique to, whiles FDs are parallel with the
initial director); moreover their flashes occur subsequently with a time
separation, though in the same (and each) half period of driving.
These scenarios are similar to those reported recently
\cite{Eber2012,May2008} for the nematic mixtures Phase 5 and Phase
4.

Electric current measurements carried out simultaneously to
pattern recording indicated strongly nonlinear current responses:
the time dependence of the current showed sharp peaks after each
polarity reversal of the applied voltage. The current nonlinearity
in 1OO8 was much more pronounced than in Phase 5. This behaviour
is attributed to the ionic conductivity of the liquid crystal. The
transient current may be due to the motion of ions during building
up a Debye screening layer at the electrodes, while the
(insulating) polyimide coating ensuring the planar alignment
blocks the charge transfer through the electrodes.

We found that, interestingly, the time instant of the flashing EC
patterns (the time of the EC contrast peak) and that of the
electric current peak coincide. This coincidence holds for all
voltages, frequencies and temperatures that we have tested. The
shape of the current signal is not affected by the occurrence of
EC significantly, indicating that it originates from the more
robust ionic effects described above. This is also supported by
the fact, that the current peaks could be observed below as well
as above the EC threshold, and even in the isotropic phase. We
think, that the current peak has a significant effect on
the formation of EC, but not vice versa; the appearance of the EC flashes is
synchronized to the current peaks. Recently we reported a
comparison \cite{Eber2012} between the measured and the
theoretically calculated time instant of the EC flashes for Phase
5. It indicated that in the experiment at ultralow $f$ EC occurred
earlier within the half period than expected from the extended
standard model of EC \cite{Krekhov2008,Krekhov2011}. We suggest
that the phase locking of EC to the ionic current peaks might be the
reason for this mismatch (the extended standard model does not
consider ionic effects). We guess that an adequate extension of
the theory to weak electrolytes could reveal this problem and
additionally explain the role of the robust current peaks in the
pattern formation; proving that, however, represents a great
theoretical challenge for the future.

Studying the threshold characteristics of the patterns we found
that the behaviour of EC and FD are essentially different.
Flexodomains have a sharp threshold, i.e. the pattern contrast
increases suddenly for $U>U_c$. For EC this holds only at high
$f$; reducing the frequency the EC threshold becomes gradually
less sharp (the contrast changes smoothly with the voltage). On
the one hand it hinders the precise determination of the EC
threshold. On the other hand, we showed that this tendency can be
followed quantitatively using an imperfect bifurcation model. In
this approach the amount of imperfection increases as the
frequency is lowered.

EC and FD have different frequency dependence of their thresholds.
At high $f$ the EC threshold is lower, while at DC driving
flexodomains are seen. Therefore it is not surprising that there
is a crossover between EC and FD at around 60 mHz, where their
thresholds become equal. Such a scenario was already anticipated
from measurements on Phase 5, but could first be demonstrated
explicitly now on 1OO8.

Interestingly, the two kinds of patterns can appear in the same
half period in some frequency range on both sides of the crossover
point, including frequencies where the two thresholds are quite
different. This is made possible by the narrow time interval and
time separation of the flashes.

The $q_{cEC}(f)$ curve of 1OO8 shows a discontinuity at $f_c
\approx 7$ Hz, indicating a crossover from conductive to
dielectric convection rolls. Interestingly, unlike similar
crossovers reported at high frequencies in other compounds, here
both the conductive and the dielectric rolls are oblique around
this crossover frequency; consequently the Lifshitz-frequency is
located in the dielectric regime. Though oblique dielectric rolls
have already been reported recently in Phase 4 (which had no
conductive regime at all) \cite{May2008}, to our knowledge 1OO8 is
the first substance which exhibits the transition from oblique
conductive to oblique dielectric rolls with increasing the
frequency of the ac voltage. The low $f_c$ indicates a fairly low
electrical conductivity which also helps distinguishing between EC
and FD patterns by increasing their time separation and may also
be responsible for the enhanced nonlinearity of the current.

Measuring the critical wave number of the flexoelectric domains
offers a way to calculate the combination $|e_{1}-e_{3}|$ of the
flexoelectric coefficients using theoretical models based either
on the one-elastic-constant approximation or on a rigourous
handling of anisotropic elasticity. It has turned out that the
values determined by the two methods differ only by about 7\%. The
reason for this small difference is that the relevant material
parameters ($K_{11}$, $K_{22}$ and $\varepsilon_a$) of 1OO8 fall
into that range, where $q_{cFD}$ is only slightly sensitive to the
elastic anisotropy. The threshold voltages of FDs, calculated from
the theoretical models using the above values of
$|e_{1}-e_{3}|$, show a satisfactory agreement with the
measured data; this proves the consistency of the models.

In cooling 1OO8 has a nematic temperature range of about 25
degrees. The temperature dependence of the elastic moduli, the
dielectric and the magnetic anisotropies was determined for the
whole nematic range. For $|e_{1}-e_{3}|$ data could be obtained
only for the lower temperature part of the nematic phase as
flexodomains did not exist for $T-T_{NI}>-8$ $^{\circ}$C. The
temperature dependence of $|e_{1}-e_{3}|$ was compared with that
of $\chi_a$; the latter being proportional to $S(T)$. It was found
that $|e_{1}-e_{3}| \propto S^2$ is satisfied, as it is expected from
the molecular theory of dipolar flexoelectricity, and also the proportionality constant was determined.

\section*{ACKNOWLEDGEMENTS}
Financial support by the Hungarian Research Fund OTKA K81250 is
gratefully acknowledged. We also thank Werner Pesch for fruitful discussions.

\end{document}